\documentclass[prb,aps,twocolumn,superscriptaddress,showpacs,preprintnumbers,amsmath,amssymb]{revtex4}
\usepackage{color}
\usepackage{graphicx}
\usepackage{ulem}

\def\textbf#1{\boldsymbol{#1}}

\begin{document}

\title{Raman Scattering Study of the Lattice Dynamics of Superconducting LiFeAs}

\author{Y. J.~Um}
\affiliation{Max-Planck-Institut~f\"{u}r~Festk\"{o}rperforschung,
Heisenbergstr.~1, D-70569 Stuttgart, Germany}

\author{J. T.~Park}
\affiliation{Max-Planck-Institut~f\"{u}r~Festk\"{o}rperforschung,
Heisenbergstr.~1, D-70569 Stuttgart, Germany}

\author{B. H. Min}
\affiliation{Department of Physics, Sungkyunkwan University,
Suwon, Gyeonggi-Do 440-746, Republic of Korea}

\author{Y. J. Song}
\affiliation{Department of Physics, Sungkyunkwan University,
Suwon, Gyeonggi-Do 440-746, Republic of Korea}

\author{Y. S.~Kwon}
\affiliation{Department of Physics, Sungkyunkwan University,
Suwon, Gyeonggi-Do 440-746, Republic of Korea}

\author{B.~Keimer}
\affiliation{Max-Planck-Institut~f\"{u}r~Festk\"{o}rperforschung,
Heisenbergstr.~1, D-70569 Stuttgart, Germany}
\author{M.~Le Tacon}
\affiliation{Max-Planck-Institut~f\"{u}r~Festk\"{o}rperforschung,
Heisenbergstr.~1, D-70569 Stuttgart, Germany}

\date{\today}

\begin{abstract}
We report an investigation of the lattice dynamical properties of LiFeAs using inelastic light scattering. Five out of the six expected phonon modes are observed. The temperature evolution of their frequencies and linewidths is in good agreement with an anharmonic-decay model. We find no evidence for substantial electron-phonon coupling, and no superconductivity-induced phonon anomalies.
\end{abstract}


\maketitle

\section{Introduction}
Following the recent discovery of superconductivity in F-doped LaFeAsO with superconducting transition temperature $T_c \sim$ 26 K,~\cite{Kamihara08} several families of FeAs-based superconductors including $RE$FeAs(O$_{1-x}$F$_{x}$) (1111-family, $RE$ = rare earth), $M$Fe$_{2}$As$_{2}$ (122-family, $M$ = Ba, Ca, Sr, K, Cs ...), $M$FeAs (111-family, $M$ = Li, Na)~\cite{Takahashi08,Rotter08,Tapp08} have been found and investigated.
All these compounds share a similar tetragonal structure based on FeAs layers. In the 1111 and 122 compounds, superconductivity emerges when the structural and spin density wave transitions in the stoichiometric parent compounds are suppressed by chemical doping or pressure.
LiFeAs deserves special attention, because it shows neither structural nor magnetic phase transitions, and superconductivity with $T_c \sim 18$ K is present at ambient pressure without any doping.~\cite{Tapp08, Wang_SSC2008, Chu_PC2009, Li_PRB2009}
The origin of superconductivity in this compound is still controversial.

\begin{figure*}[t]
\includegraphics[width=0.9\linewidth]{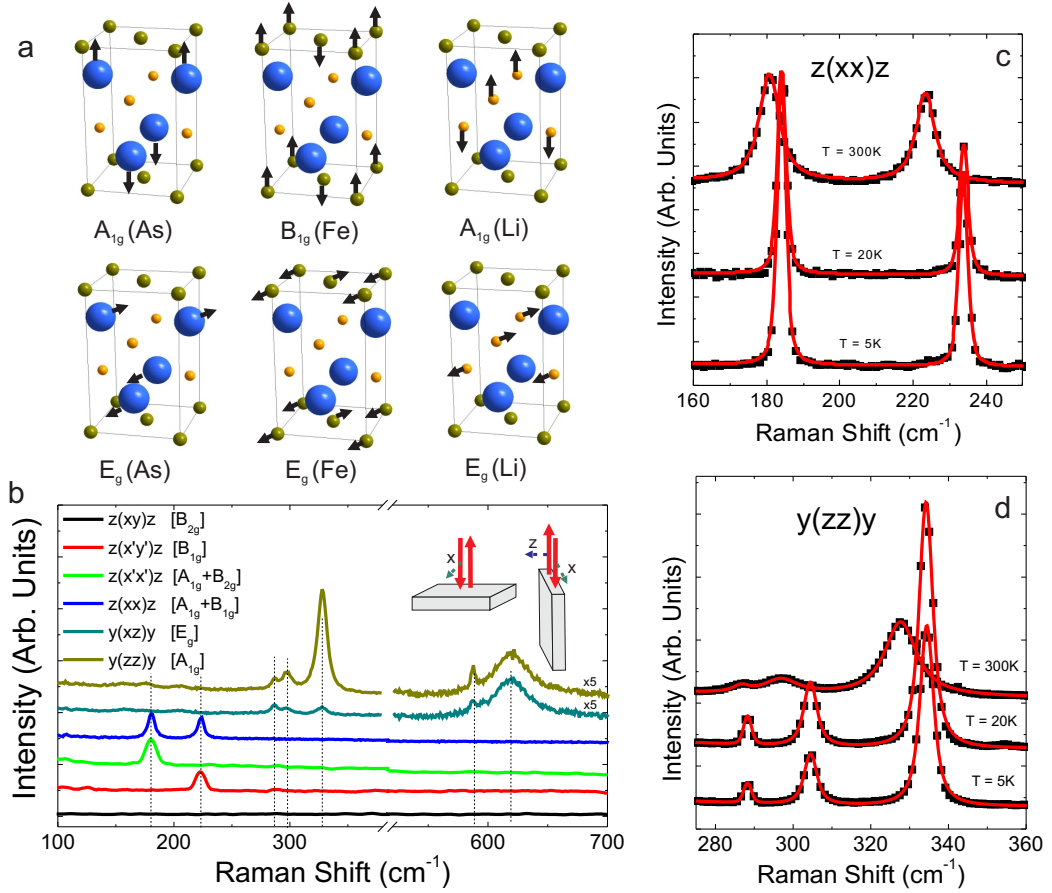}
\caption{(Color online) a) Atomic displacement patterns of the Raman-active optical modes of LiFeAs. b) Room temperature Raman spectra in $z(xy)z$, $z(x^{\prime}y^{\prime})z$, $z(x^{\prime}x^{\prime})z$, $z(xx)z$, $y(xz)y$, and $y(zz)y$ configurations (note that the spectra in $z(x^{\prime}y^{\prime})z$ and $z(x^{\prime}x^{\prime})z$ configurations have been recorded in a lower-resolution mode to maximize the signal). Spectra have been shifted vertically for clarity. Starting from the left, the peaks are assigned to A$_{1g}$(As), B$_{1g}$(Fe), E$_{g}$(1), E$_{g}$(2), and A$_{1g}$(Li) vibrations. c) A$_{1g}$(As) and B$_{1g}$(Fe) phonons for selected temperatures (room temperature, $T \sim T_{c}$, and the base temperature of our cryostat). Black squares are the data, the red line is the result of a fit following the procedure described in the text. The spectra have been shifted vertically for clarity. d) Same plot for E$_{g}$(1), E$_{g}$(2), and A$_{1g}$(Li) phonons.}
\label{Fig1}
\end{figure*}

The presence of weak local moments~\cite{Zhang_PRB2009} and normal-state antiferromagnetic (AF) fluctuations,~\cite{Jeglic_PRB2010, Wang_PRB2011, Taylor_PRB2011, Qureshi_arxiv} combined with prediction of weak electron-phonon coupling~\cite{Jishi_ACMP2010} seem to support an unconventional pairing mechanism, as in other FeAs compounds. 
On the other hand, London penetration depth and small-angle neutron scattering,~\cite{Inosov_PRL2010} angle-resolved photoemission spectroscopy (ARPES),~\cite{Borisenko_PRL2010} microwave surface impedance~\cite{Imai_JPSJ2011} and several other measurements~\cite{Inosov_PRB2011} indicate a superconducting gap ratio 2$\Delta \sim$ 4 $k_BT_c$ closer to the weak-coupling limit than other Fe-based superconductors. In addition, ARPES measurements have been interpreted as evidence of strong electron-phonon coupling,~\cite{Kordyuk_PRB2011} possibly enhanced by the AF spin fluctuations.~\cite{Huang_PRB2010, Li_APL2011}
Taken together, these data rather suggest conventional electron-phonon driven pairing in this compound. The present study is motivated by the absence of experimental investigations of the lattice dynamics that would allow a direct determination of the electron-phonon coupling strength in LiFeAs.


\begin{table*}
\caption{Calculated Raman-active phonon frequencies and selection rules from Refs.~\onlinecite{Jishi_ACMP2010} and~\onlinecite{Huang_PRB2010} and comparison to our experimental results (see text for the definition of the parameters).}
\begin{ruledtabular}
\begin{tabular*}{0.95\columnwidth}{c|c|c|c|c|c|c|c|c}
 Mode     & Polarization & Selection Rule & Calc. Freq. (cm$^{-1}$) &  Calc. Freq. (cm$^{-1}$) & \multicolumn{4}{c}{ Exp. fitting Parameters (this work)} \\
(atom)    &  &  & from Ref.~\onlinecite{Jishi_ACMP2010} & from Ref. ~\onlinecite{Huang_PRB2010} &  $\omega_0$ (cm$^{-1}$)      &  $C$ (cm$^{-1}$)&  $\Gamma_0$ (cm$^{-1}$) & $\Gamma$ (cm$^{-1}$) \\
\hline
E$_{g}$ (As) & in-plane & xz &121 cm$^{-1}$& 173.7 & - & - & - & -\\
A$_{1g}$ (As) & c-axis & xx, zz, x$^\prime$x$^\prime$ &188 & 183.3 & 185.1&  1 &  0.24 & 1.3 \\
B$_{1g}$ (Fe) &c-axis & x$^\prime$y$^\prime$, xx&225 & 207.5 & 237.8&  4 &  0.07 & 1.6 \\
E$_{g}$ (Fe) & in-plane & xz &240 & 224.7 & 289.2&  0.9 &  0 & 1.1 \\
E$_{g}$ (Li) & in-plane & xz &294 & 233 & 308.2 &  3.7 &  0 & 3.3 \\
A$_{1g}$ (Li) &c-axis & xx, zz, x$^\prime$x$^\prime$ &356 & 299.8 & 338.2&  4 & 0 & 3.5 \\
\end{tabular*}
\end{ruledtabular}
\label{Table1}
\end{table*}

\section{Experimental Details}

Single crystals of LiFeAs with $T_c \sim 18$ K were grown in a sealed tungsten crucible using the Bridgman method.~\cite{Song_APL2010}
Since LiFeAs crystals are extremely air sensitive, they were cleaved and mounted on the cold finger of a helium-flow cryostat in a glove box under Ar atmosphere.
The Raman spectra were taken in backscattering geometry in a JobinYvon LabRam 1800 single-grating spectrometer equipped with a razor-edge filter and a Peltier-cooled CCD camera.
We used a linearly polarized He$^{+}$/Ne$^{+}$ mixed-gas laser with a wavelength of 632.817 nm for excitation.
The laser beam was focused through a 50$\times$ microscope objective to a $\sim 5$ $\mu$m diameter spot on the sample surface. The power of the incident laser was kept less than 2~mW to avoid laser-induced heating. In order to determine the precise phonon frequencies at every temperature, Ne emission lines were recorded between all measurements. For data analysis, all phonon peaks were fitted by Lorentzian profiles, convoluted with the spectrometer resolution function (a Gaussian line of 2 cm$^{-1}$ full width at half maximum (FWHM)).


\section{Experimental Results}
\subsection{Mode Assignment}

The lattice symmetry of LiFeAs is described by the space group $P4/nmm$ ($D_{4h}^7$), with the Li, Fe and As atoms located at the 2c, 2b and 2c Wyckoff positions, respectively.
From group symmetry analysis, 15 zone-center optical phonons are expected, among which 2A$_{1g}$, 1 B$_{1g}$ and 3E$_g$ modes are Raman active.~\cite{Rousseau_JRS1981} The corresponding atomic displacement patterns are sketched in Fig.~\ref{Fig1}-a and the Raman selection rules as well as the frequencies calculated within density functional theory~\cite{Jishi_ACMP2010, Huang_PRB2010} are listed in Table 1.
In Fig.~\ref{Fig1}-b, we show the Raman spectra measured at room temperature for several scattering geometries with incident light wave vectors along the c-axis ($z(xy)z$, $z(x^{\prime}y^{\prime})z$, $z(x^{\prime}x^{\prime})z$, $z(xx)z$ configurations in Porto notation) or along the y axis ($y(zz)y$ and $y(xz)y$).
As expected, no phonon modes are observed in the $z(xy)z$ geometry which selects the B$_{2g}$ symmetry.

The measurements in the $z(x^{\prime}y^{\prime})z$, $z(x^{\prime}x^{\prime})z$, $z(xx)z$ configurations select phonons in the B$_{1g}$, A$_{1g}$ and A$_{1g}$+B$_{1g}$ symmetries, respectively, which allows us to unambiguously assign the modes at 181 cm$^{-1}$ and 223 cm$^{-1}$ to the A$_{1g}$(As) and B$_{1g}$(Fe) c-axis polarized vibrations of the FeAs planes. These frequencies are in better agreement with the calculations of Ref.~\onlinecite{Jishi_ACMP2010}, which are based on the experimental lattice constants and atomic coordinates, than with those of Ref.~\onlinecite{Huang_PRB2010}, which were obtained using the relaxed ones. The mode frequencies are also very close to those of analogous phonons in the 122~\cite{Choi_PRB2008, Litvinchuk_PRB2008, Rahlenbeck09, Chauviere_PRB2009} or 1111 compounds.~\cite{Hadjiev_PRB2008, LeTacon_PRB2008, Gallais_PRB2008, Zhang2_PRB2009} We note that the second A$_{1g}$ mode, involving mainly motions of the Li atoms along the c-axis, is not observed here.

For measurements with light polarization in the ac plane, we observe three modes at 287 cm$^{-1}$, 297 cm$^{-1}$ and 328 cm$^{-1}$, as well as two other features at high energy around 585 and 620 cm$^{-1}$.
In the $y(xz)y$ geometry, only the in-plane E$_{g}$ modes are expected. The 328 cm$^{-1}$ phonon is very weak in this configuration, but becomes much more intense than the two others upon switching to the $y(zz)y$ configuration, and should therefore be attributed to the second A$_{1g}$ mode rather than to an in-plane E$_{g}$ phonon. The proximity of this mode frequency with the calculation for the A$_{1g}$(Li) mode further confirms this assignment.
The two remaining modes at 287 cm$^{-1}$ and 297 cm$^{-1}$ are finally attributed to E$_{g}$ phonons, despite significant disagreement with the calculated frequencies (240 (225) and 290 (233) cm$^{-1}$ from Ref.~\onlinecite{Jishi_ACMP2010} and \onlinecite{Huang_PRB2010}, respectively).
We note that the selection rules are not perfectly respected here as, for instance, the E$_{g}$ modes are still visible in the $y(zz)y$ geometry.
The ac plane is not a good cleavage plane, giving rise to a rough surface hard to align accurately in the glove box. As a consequence we always get a slight misalignment in the crystal that causes the observed mode leakages.
The lowest-energy mode predicted by these calculations could not be observed in this study.
It is worth noting that the two A$_{1g}$ modes are not visible in the same scattering geometry. In z(xx)z or z(x$^{\prime\prime}$x')z configurations, the phonon intensity is proportional $(2\alpha_{xx})^2$, and to $(\alpha_{zz})^2$ in the y(zz)y configuration, where $\alpha_{zz}$ and $\alpha_{xx}$ are the diagonal elements of the $A_{1g}$ Raman tensor. Our results imply that the Raman tensor components for the two A$_{1g}$ phonons are different: we have indeed $\alpha_{zz} \sim 0 << \alpha_{xx}$ for the A$_{1g}$ As mode while we have $\alpha_{xx} \sim 0 << \alpha_{zz}$ for the A$_{1g}$ Li mode.
Such anisotropy of the Raman tensor has for instance been reported in the case of SrFe$_2$As$_2$,~\cite{Litvinchuk_PRB2008} where $\alpha_{xx} \sim 0 << \alpha_{zz}$ has been observed for the A$_{1g}$ mode.

\begin{figure}
\includegraphics[width=0.8\linewidth]{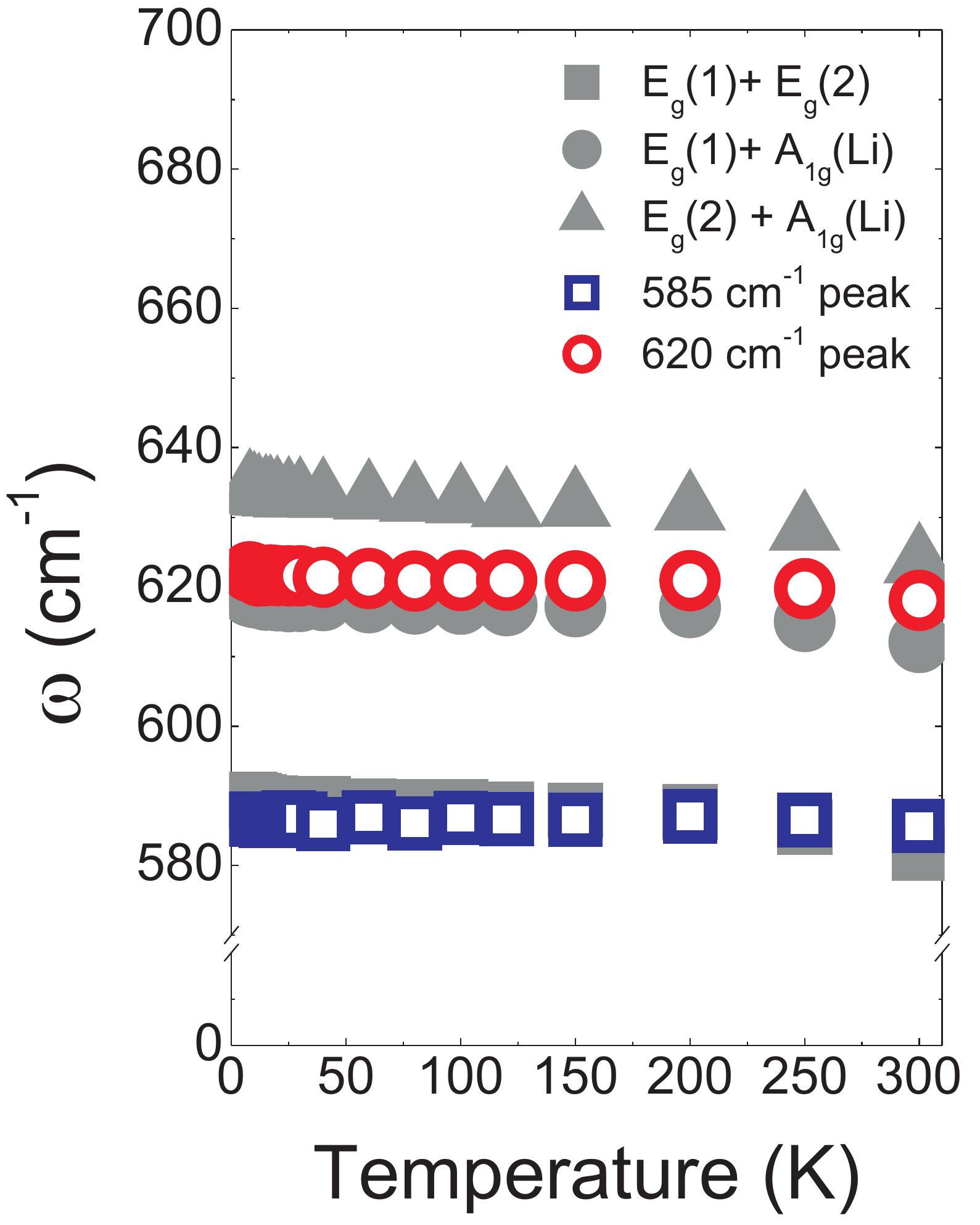}
\caption{(Color online) Comparison of the temperature dependence of the frequencies of the high-energy modes at 585 and 620 cm$^{-1}$ with the sums of the frequencies of modes observed in $y(zz)y$ and $y(xz)y$ polarizations.}
\label{Fig2}
\end{figure}

The energy of the two features observed at 585 and 620 cm$^{-1}$ is much higher than the highest calculated phonon energy in this system (which is about 300 cm$^{-1}$~\cite{Huang_PRB2010, Jishi_ACMP2010}). They can therefore not be attributed to single-phonon modes, but rather appear to be due to two-phonon scattering.
This is further confirmed by the observation that 585 cm$^{-1}$ is almost exactly equal to the sum of the frequencies of the
two E$_g$ modes at all temperatures, while the frequency of the second feature is always located between the sums of the energy of the A$_{1g}$(Li) phonon with each of the two E$_g$ modes (Fig.~\ref{Fig2}). We note that the large width of this feature compared to the one at 585 $cm^{-1}$ ($\sim 35$ cm$^{-1}$ vs. $\sim 3.5$ cm$^{-1}$) indeed suggests that it can be further decomposed into several features that are not individually resolved. This is not clear why these features are observed only when the light is polarized in the ac plane, and further investigation are required to understand this point.


\begin{figure}
\includegraphics[width=0.9\linewidth]{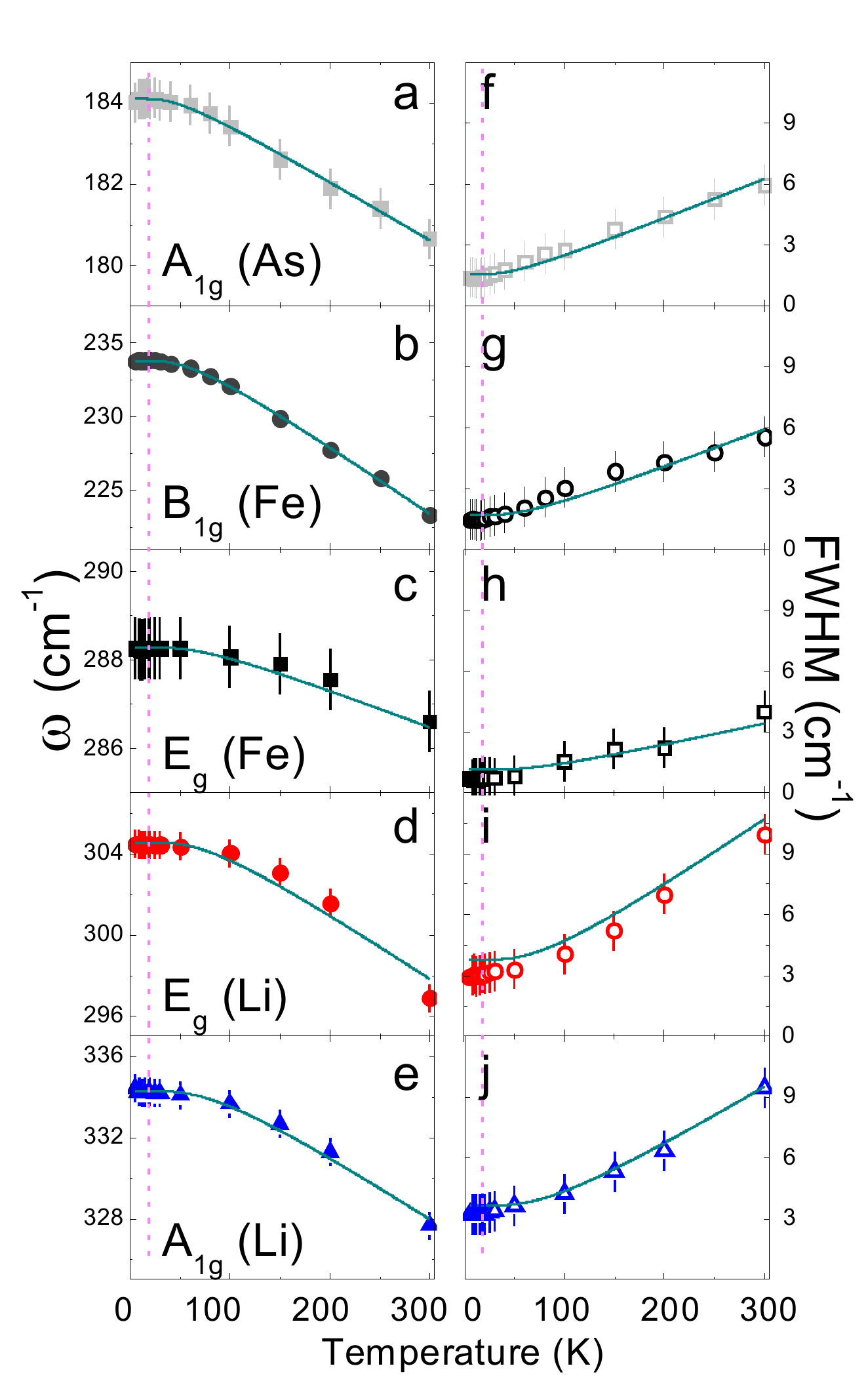}
\caption{(Color online) Temperature dependence of frequencies (panels a-e) and linewidths (panels f-j) of the five observed Raman-active modes. The pink dashed line marks $T_c$, and the green line is the result of a conventional phonon anharmonic model (see text).}
\label{Fig3}
\end{figure}

As seen in Figs.~\ref{Fig1}-c and d, all of these phonons are found to harden and sharpen as the temperature decreases. Note that the intensity drop seen between 20 and 5 K in Fig.~\ref{Fig1}-d was not reproducible and shall not be considered as a superconductivity-induced effect. Phonon intensities will not be further considered.
Figure~\ref{Fig3} shows the temperature dependence of their positions and linewidths. The absence of phonon anomalies at any temperature (such as the splitting of the E$_g$ modes predicted~\cite{Huang_PRB2010} and observed~\cite{Chauviere_PRB2009} at the structural phase transition in BaFe$_2$As$_2$) confirms that LiFeAs does not undergo any structural or magnetic phase transitions. Figure~\ref{Fig3} also shows the results of fits of the experimental data points to a simple expression for the temperature dependence of the frequency and linewidth of an optical phonon resulting from symmetric anharmonic decay, \textit{i.e.} decay into two acoustic modes with identical frequencies and opposite momenta:~\cite{Klemens66, Menendez84}
\begin{equation}\label{e1}
\omega_{ph}(\textit{T}) = \omega_{0} - C\bigg[1+\frac{2}{e^{\frac{\hbar\omega_{0}}{2k_{B}T}}-1}\bigg]
\end{equation}
\begin{equation}\label{e2}
\Gamma_{ph}(\textit{T}) = \Gamma_{0} + \Gamma\bigg[1+\frac{2}{e^{\frac{\hbar\omega_{0}}{2k_{B}T}}-1}\bigg]
\end{equation}

where  $C$ and $\Gamma$ are positive constants, $\omega_{0}$ is the bare phonon frequency, and $\Gamma_{0}$ a residual (temperature independent) linewidth originating from sample imperfections or electron-phonon interactions.
We find excellent agreement with this expression. The fitting parameters are summarized in Table 2.

In all cases, the residual width $\Gamma_{0}$ is found to be vanishingly small, which confirms the high quality of our crystals.
The temperature dependent coefficient $\Gamma$ is always much larger than the residual width, indicating that lattice anharmonicity is the principal source of decay for the phonons in LiFeAs, in clear contrast with the recent observations of Litvinchuk \textit{et al.} in Pr$_x$Ca$_{1-x}$Fe$_2$As$_2$.~\cite{Litvinchuk_PRB2011}
Three kinks were recently observed around 15, 30 and 44 meV (121, 242 and 355 cm$^{-1}$) in energy distributions curves measured by ARPES on LiFeAs, and attributed to electron-phonon coupling.~\cite{Kordyuk_PRB2011}
Our present data cannot completely rule out the possibility of such a coupling, but excludes it to occur at the zone center where our measurements were carried out. To confirm the electron-phonon coupling origin of this kink, a study of the phonon dispersion is this compound is therefore required~\cite{Letacon_PRB2009, Hanh_PRB2009}.

No changes in any of the phonon frequencies or linewidths are observed at the superconducting transition, as previously reported for K-doped BaFe$_2$As$_2$~\cite{Rahlenbeck09} and F-doped NdFeAsO.~\cite{Gallais_PRB2008} This is consistent with the relatively small amplitude of the superconducting gap (2$\Delta \sim$ 4.0 $k_BT_c \sim$ 50 cm$^{-1}$ compared to the phonon frequencies reported here.~\cite{Inosov_PRL2010} To our knowledge, the only FeAs-based compound in which superconductivity-induced changes in the Raman phonons are clearly observed is Pr$_x$Ca$_{1-x}$Fe$_2$As$_2$.~\cite{Litvinchuk_PRB2011}

\section{Conclusions}
In conclusion, we have reported a lattice dynamical study of superconducting LiFeAs. We observed five modes out the six expected. Their frequencies are in good overall agreement with values predicted by density functional calculations, and their temperature dependence is well described by an anharmonic-decay model. We observed neither clear anomalies associated with the superconducting transition nor any evidence for substantial electron-phonon coupling. 

\nonumber\section{Acknowledgements}
We thank A. Schulz for technical support and D. Inosov, J. K\"{o}hler and A. Bussmann-Holder for useful suggestions. This work has been supported by the European project SOPRANO (Grant No. PITN-GA-2008-214040), by the Basic Science Research Program (2010-0007487), the Mid-career Researcher Program (2010-0029136) and the Nuclear R\&D Programs (2006-2002165) through NRF funded by the Ministry of Education, Science and Technology of Korea.


\end{document}